\documentclass[%
 reprint,
 amsmath,amssymb,
 aps,
]{revtex4-2}

\usepackage{graphicx}
\usepackage{dcolumn}
\usepackage{bm}
\usepackage{hyperref}
\usepackage[mathlines]{lineno}

\usepackage{lineno}
\usepackage{xcolor}
\usepackage{booktabs}

\usepackage{braket}
\usepackage{amsmath,mleftright}
\usepackage{latexsym,amssymb}
\usepackage{amsmath}
\usepackage{amstext}
\usepackage{mathrsfs}
\usepackage{amsfonts}
\usepackage{subcaption}

\newtheorem{definition}{Definition}[section]
\newtheorem{lemma}{Lemma}[section]
\def\nn{\nonumber}
\def\pa{\partial}
\def\l{\left}
\def\r{\right}

\def\l{\left}
\def\r{\right}
\def\a{\alpha}

\def\c{\gamma}

\def\m{\mu}
\def\n{\nu}

\def\la{\lambda}

\def\G{\Gamma}

\def\bs{\boldsymbol}
\def\rm{\mathrm}

\newcommand{\ie}{\text{i.e.~}}

\def\tr{\mathrm{tr}}

\definecolor{JNcolor}{rgb}{0.9,0.1,0.1}

\definecolor{ABcolor}{rgb}{0,0,1}

\begin{document}

\title{Revisiting holographic codes with fractal-like boundary erasures}

\author{Abhik Bhattacharjee}
\email{abhikbhattacharjee@hri.res.in}
\affiliation{Harish–Chandra Research Institute, A CI of Homi Bhabha National Institute, Chhatnag Road, Jhunsi, Prayagraj, Uttar Pradesh 211019, India}

\author{Joydeep Naskar}
\email{naskar.j@northeastern.edu}
\affiliation{Department of Physics, Northeastern University, Boston, MA 02115, USA}
\affiliation{The NSF AI Institute for Artificial Intelligence and Fundamental Interactions, Cambridge, MA, U.S.A.}

\date{\today}
             
\begin{abstract}
In this paper we investigate the code properties of holographic fractal geometries initiated in \cite{Pastawski:2016qrs}. We study reconstruction wedges in $AdS_3/CFT_2$ for black hole backgrounds, which are in qualitative agreement with the vacuum-AdS approximation using generalized entanglement entropy in \cite{Bao:2022tgv}. In higher dimensions, we study reconstruction wedges for the infinite, straight strip in $AdS_{d+1}/CFT_{d}$ and clarify the roles of `straight' and `infinite' in their code properties. Lastly, we comment on uberholography from the perspective of complexity transfer and one-shot holography.
\end{abstract}

\maketitle


\section{\label{sec:level1}Introduction}
The study of AdS/CFT correspondence\cite{Maldacena:1997re} from the lens of quantum information theory has produced several interesting results in the last decade. Influential ideas about quantum mechanics and gravity, such as the AMPS\cite{Almheiri:2012rt}, ER$=$EPR\cite{Maldacena:2013xja}, complexity\cite{Susskind-complexity,Susskind:2018pmk}, gravity from entanglement\cite{Faulkner:2013ica}, to name a few, posed interesting questions and answers. A pivotal point was the seminal idea of recognizing the bulk-boundary correspondence as a quantum error-correcting code (QECC)\cite{Verlinde:2012cy,Almheiri:2014lwa,Pastawski:2015qua}. This program has formalized the idea of bulk reconstruction and improved our understanding profoundly, say for example, resolution of the commutator puzzle and the notion of subregion duality\cite{Jafferis:2015del, Dong:2016eik,Hayden:2016cfa}. Several mysteries related to the blackhole information paradox have been uncovered as well with many different proposals\cite{Papadodimas:2012aq,Almheiri:2019psf,Akers:2022qdl,Bueller:2024zvz,Raju:2020smc}. Purely mathematical ideas of operator algebra, found close relations to gravity and quantum field theories\cite{Witten:2018zxz,Harlow:2016vwg,Pollack:2021yij}. See the reviews \cite{Harlow:2018fse,Kibe:2021gtw,Chen_2022} that capture some of the essence of this program. While holography proved to be a theoretical laboratory for ideas from quantum information, there is also considerable interest in how holographic studies of quantum error correction can influence quantum error correction as applied in quantum computing. In our current work, however, we restrict ourselves to the former, i.e., using ideas and tools from quantum information theory and applying them to holography. One central idea that we borrow is that of quantum error correction.

In Operator Algebra Quantum Error Correction (OAQEC), we consider a code subspace $\mathcal H_C=P\mathcal H$ of a larger physical Hilbert space $\mathcal H$, where $P$ denotes the orthogonal projector from $\mathcal H$ to $\mathcal{H}_C$. Given a code subspace $\mathcal H_C$ and a logical subalgebra $\mathcal A$, if a subsystem $R$ of $\mathcal H$ is \emph{correctable} with respect to $\mathcal A$, then $\mathcal A$ can be reconstructed on $R^c$. We call a subsystem $R$ of $\mathcal H$ correctable with respect to $\mathcal A$ iff $[PYP,X]=0$ for all $X\in\mathcal A$ and every operator $Y$ supported on $R$.

For holographic codes, the correctability condition is given by the entanglement wedge hypothesis. Let us define the entanglement wedge as follows.
\begin{definition}[Entanglement wedge]
The entanglement wedge of a boundary region $R\subseteq \pa B$ is the bulk region $\mathcal E[R]\subseteq B$, whose boundary is $\pa\mathcal E[R]=\chi_R\cup R$, $\chi_R$ being the minimal surface for $R$, \ie the co-dimension-one surface in $B$ separating $R$ from its boundary complement $R^c$.    
\end{definition}
We use the above definition of entanglement wedge to make the following proposition about quantum error correction properties of holographic codes.
\begin{definition}[Entanglement wedge hypothesis]
If a bulk point $x$ is contained in the entanglement wedge $\mathcal E[R]$ of boundary region $R\subseteq \pa B$, then $R^c\subseteq\pa B$ is correctable with respect to the bulk logical subalgebra $\mathcal A_x$.    
\end{definition}

In general, entanglement wedge reconstruction can be state-dependent by prescription of the quantum extremal surfaces\cite{Engelhardt:2014gca, Akers:2021fut}. It was observed in \cite{Hayden:2018khn,Akers:2019wxj} that there is a breakdown of entanglement wedge when the bulk contains a mixed state whose entropy is large enough. As a result, the entanglement wedge hypothesis does not hold. Instead, one defines a reconstruction wedge and the condition for error correctability is upgraded to the containment of $x$ in the reconstruction wedge. We define the reconstruction wedge below.
\begin{definition}[Reconstruction wedge]
The \textit{reconstruction wedge} of a boundary region A is the intersection of all entanglement wedges of A for every state in the code subspace, pure or mixed. It is the region of space-time within which bulk operators are guaranteed to be reconstructible from the boundary in a state-independent manner.    
\end{definition}

The idea of uberholography was first introduced in \cite{Pastawski:2016qrs}. Uberholography is a property exhibited by holographic codes defined on bulk manifolds with asymptotically negative curvature, allowing the bulk logical algebra to be supported on a boundary region with a fractal structure. Uberholography suggests that higher-dimensional bulk geometry can emerge from lower-dimensional systems with non-integer Hausdorff dimensions, providing insights into the relationship between fractal geometry and quantum error correction. See \cite{Pastawski:2016qrs,Gesteau:2020hoz,Chen:2019uhq,Bao:2022vxc,Ageev:2022awq,Bao:2022tgv,Wang:2024gvz} for related works on uberholography.

In this work, we fill up some gaps in the uberholography literature. For example, uberholography in $AdS_3/CFT_2$ was considered in \cite{Bao:2022tgv} in the vacuum-AdS approximation neglecting backreaction, whereas a BTZ blackhole background with just the first level of erasure on the boundary was considered in \cite{Ageev:2022awq}. We consider the BTZ blackhole metric for an arbitrary level $m$ of erasures on the boundary and plot the parameter $r$ at criticality. Moreover, the higher dimensional Cantor-set like erasures in $AdS_{d+1}/CFT_{d}$ were considered upto level 1 and all arguments were made using the entanglement wedge hypothesis\cite{Ageev:2022awq}. We upgrade those arguments for generic level $m$ using the reconstruction wedge, taking the bulk entanglement entropy into account. We shed light on the geometrical properties of the infinitely long straight strip, especially how `straightness' and `infinite' length play crucial role in their code properties. Lastly, we comment on uberholography from perspective of complexity transfer\cite{Engelhardt:2023xer} and one-shot holography\cite{Akers:2020pmf,Akers:2023fqr}, incorporating uberholography in these modern paradigms consistently.

The organization of the paper is as follows: In Section \ref{sec: review}, we give a brief review of uberholography in $AdS_3/CFT_2$, in section \ref{sec:BTZ-AdS3} we consider a BTZ blackhole geometry in $AdS_3$ and establish the qualitative agreement with the results obtained considering generalized entropy in the vacuum-$AdS_3$ approximation. In section \ref{sec:d_uber}, we consider higher-dimensional generalizations of uberholography for Cantor-set-like erasures. Lastly, our comments on complexity transfer and one-shot state merging is in section \ref{sec:comments}.

\section{Review of uberholography}\label{sec: review}
In the $AdS/CFT$ correspondence, the Ryu-Takayanagi formula relates the entanglement entropy of a subregion $A$ in the boundary CFT to the area of the minimal surface in the bulk homologous to $A$,
\begin{equation}\label{eq:RTformula}
    S(A)=\frac{\rm{Area} (\mathcal{\chi}_A)}{4G_N},
\end{equation}
where $S(A)$ is the boundary CFT entanglement entropy of the subregion $A$ and $\mathcal{\chi}_A$ is the Ryu-Takayanagi surface for $A$ and $G_N$ is Newton's constant. In our text, entanglement wedge refers to the wedge bounded by the RT surface. This can be generalized into the quantum extremal surface by taking account of the bulk entanglement entropy $S_{bulk}$ contained within the wedge,
\begin{equation} \label{eq:qes_formula}
    S(A)=\frac{\rm{Area} (\mathcal{\tilde{\chi}}_A)}{4G_N} + S_{bulk},
\end{equation}
where $\tilde{\chi}$ is the quantum minimal surface extremizing the equation $\ref{eq:qes_formula}$ whose RHS is defined as the \emph{generalized entropy}.

We will now review uberholography in $AdS_3/CFT_2$ in vacuum $AdS_3$ background with entanglement wedge and reconstruction wedge arguments. For uberholography in higher dimensions, see \cite{Bao:2022vxc,Ageev:2022awq}.
\subsection{Uberholography in $AdS_3/CFT_2$}
\begin{figure}[h!]
    \centering
    \includegraphics[width=\linewidth]{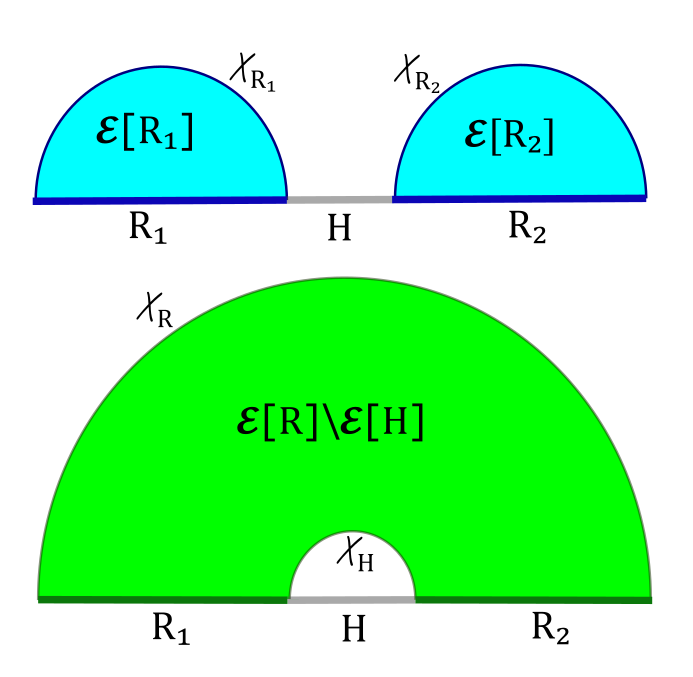}
    \caption{The two possible entanglement wedge candidates at level-1 in $AdS_3/CFT_2$: disconnected (top) and connected (bottom).}
    \label{fig:conn_disc1}
\end{figure}

The key idea is as follows: Consider a boundary region $R$ of length $|R|$. Let us punch a hole $H$ of size $(1-r)|R|$ where $0\leq r \leq 1$ symmetrically, such that the boundary is now divided into three regions $R_1$, $H$ and $R_2$ (see figure \ref{fig:conn_disc1}), such that
\begin{equation}
    |R_1|= |R_2|= \frac{r}{2}|R|, \quad |H|=(1-r)|R|,
\end{equation}
which we call the level-1 of hole punching. This hole is an erasure on the boundary.

There are two possible entanglement wedge candidates, one disconnected and one connected, depending on the value of $r$. The transfer of dominance of the two phases happens at a critical value of $r$, where the areas of the Ryu-Takayanagi surfaces of connected and disconnected wedges are equal,
\begin{equation}\label{eq:equal-areas}
    |\chi_{R_1}|+|\chi_{R_2}|=|\chi_{R}|+|\chi_{H}|,
\end{equation}
In this section from now on, when we refer to $r$, it corresponds to the critical geometry, unless otherwise stated. In $AdS_3/CFT_2$, the area of the minimal surface homologous to boundary subregion $A$ is the length of the geodesic $\chi_A$ joining the the two end points of $A$ and is given by\cite{Calabrese:2004eu,Brown:1986nw},
\begin{equation}\label{eq:minimalads3}
    |\chi_A|=2L_{AdS} \log{\left(\frac{|A|}{\epsilon}\right)},
\end{equation}
where $L_{AdS}$ is the $AdS$-radius and $\epsilon$ is the short distance cutoff. Inserting the expression \ref{eq:minimalads3} in equation \ref{eq:equal-areas},  it turns out that we have $r=2(\sqrt{2}-1)$.

Now let us punch holes $H_1$ and $H_2$ in a similar fashion in $R_1$ and $R_2$ such that
\begin{equation}
 |H_1|=|H_2|=(1-r)|R_1|=(1-r)\frac{r}{2}|R|   
\end{equation}
There are four disconnected regions remaining each of length $\left(\frac{r}{2}\right)^2 |R|$. We call this level $2$ of hole punching. Repeating this process, we can go upto an arbitrary number of level $m$ of hole punching. The connected entanglement wedge has an RT surface with area
\begin{equation}
\begin{aligned}
    |\chi_{R'}|_\rm{conn.}=2L_{AdS}\times\\
    \times\left[ \log\left(\frac{|R|}{a}\right)+
    \sum_{j=1}^m 2^{j-1} \log\left(\frac{ \left(\frac{r}{2}\right)^{j-1}(1-r)|R|}{a} \right) \right],
\end{aligned}
\end{equation}
whereas the disconnected phase has a RT surface with area,
\begin{equation}
    |\chi_{R'}|_\rm{disc.}= 2L_{AdS} \left[2^m \log\left({\frac{(\frac{r}{2})^m |R|}{a}}\right) \right].
\end{equation}
Equating these surface areas and solving for $r$, we find that $r=2(\sqrt{2}-1)$ for any arbitrary level $m$. This establishes the existence of a quantum error-correcting code in the form of uberholography on a fractal boundary geometry that is supported by a very small measure in the boundary approaching $0$ as level $m\rightarrow\infty$ and short-distance cutoff $\epsilon\rightarrow 0$. 

Let us now define the code distance of this uberholographic code. Consider the case where we are at a level $m$ such that the each small disconnected boundary subregion at this level is of length $\epsilon$. There are $2^m$ such boundary subregions. We define the distance of the code with operator algebra $\mathcal{A}$ in bulk region $X$ to be
\begin{equation}\label{codedist}
d(\mathcal A_X)\leq \frac{|R_{min}|}{a}=2^m=\left(\frac{|R|}{a}\right)^{\alpha},
\end{equation}
where
\begin{equation}
    \alpha=\frac{\log{2}}{\log{2/r}}=\frac{1}{\log_2{(\sqrt{2}+1})}=0.786,
\end{equation}
so the distance is bounded above by some $n^{\alpha}$.

\subsubsection{Reconstruction wedges in uberholography}
\begin{figure}[h!]
    \centering
    \includegraphics[width=\linewidth]{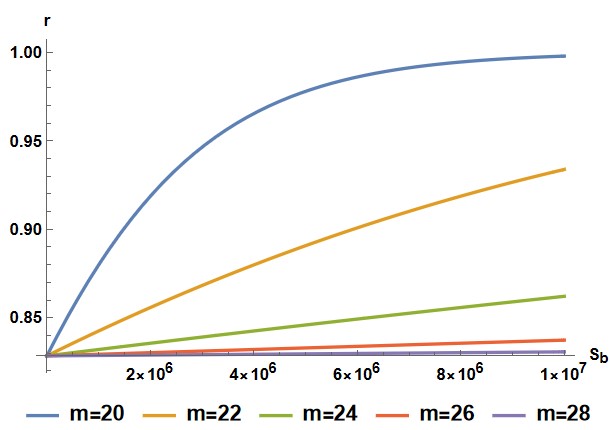}
    \caption{$r$ vs $S_b$ for different values of $m$. We see as $m\rightarrow\infty$, the curve approaches the lines $r=2(\sqrt{2}-1)$.}
    \label{fig:r-vs-Sb-diff-m}
\end{figure}

Now we introduce a bulk entanglement entropy $S_b$ in the center of the bulk. We do not explicitly describe the mechanism of how this might be achieved. We will ignore the backreaction and continue to work in the vacuum $AdS_3$ metric. With an implicit assumption at leading order, the connected phase now carries a contribution from this bulk entanglement entropy $S_b$, whereas the disconnected phase does not. At level 1, the following inequality must be satisfied,
\begin{equation}
    |\chi_{R_1}|+|\chi_{R_2}| \geq |\chi_{R}|+ |\chi_{H}| + 4G\textbf{$S_b$}
\end{equation}
hinting an upper bound on \textbf{$S_b$} for a scheme for bulk reconstruction of deep interior operators to exist. This perceived upper bound is given by
\begin{equation}
    4G\textbf{$S_b$}\leq 2L_{AdS} \log\left({\frac{(r/2)^2}{(1-r)}}\right).
    \label{leve1Sb}
\end{equation}
At a generic level $m$, we have the following constraint on the critical value of $r$,
\begin{equation}
    \frac{\left(\frac{r}{2}\right)^2}{(1-r)}=e^{4G\textbf{$S_b$}/2L_{AdS}(2^m-1)}.
\end{equation}
In the limit $m\rightarrow\infty$, the $RHS\rightarrow 1$, approaching the value of $r$ in $\textbf{$S_b$}=0$ limit,
\begin{equation}
    r\rightarrow 2(\sqrt{2}-1).
    \label{level-m-infinity-r}
\end{equation}
Therefore, we see that after introducing a bulk entanglement entropy $S_b$, $r$ is no longer independent of $m$ and depends explicitly on both $S_b$ and $m$ (see figure \ref{fig:r-vs-Sb-diff-m}).

However, when $m$ is sufficiently large enough, $r$ approaches the value it has in the absence of a bulk entanglement entropy term. This demonstrates the robustness of uberholography against the breakdown of entanglement wedge. In section \ref{sec:BTZ-AdS3}, we will drop the vacuum-AdS approximation and work explicitly in a BTZ blackhole metric.

\section{BTZ black hole background in $AdS_3$}\label{sec:BTZ-AdS3}
We will begin with the non-rotating BTZ black hole metric $AdS_3$, given by
\begin{equation}\label{eq:BTZ-metric}
    d s^{2}=\frac{L_{AdS}^{2}}{z^{2}}\left(-f(z) d t^{2}+\frac{d z^{2}}{f(z)}+dy^{2}\right),
\end{equation}
where $f(z)=1-\frac{z^{2}}{z_{h}^{2}}$ and the temperature $T$ of the dual CFT is related to the horizon radius, $T=\frac{1}{2\pi z_h}$. In this metric, the entanglement entropy of a subregion of length $l$ is given by 
\begin{equation}
     S(l)=\frac{c}{3} \log \left(\frac{\sinh \left(\pi T l \right)}{\pi T \epsilon}\right),
\end{equation}
to leading order, where $\epsilon$ is the short-distance cutoff and $G_N$ is Newton's constant.

\begin{figure}[h!]
    \centering
    \includegraphics[scale=0.6]{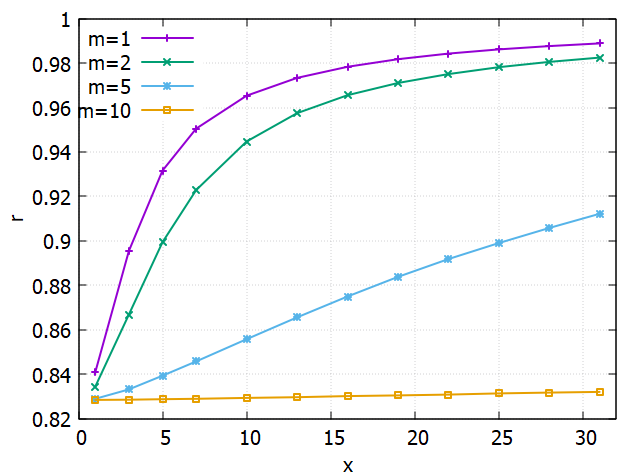}
    \caption{$r$ vs. $x$ plot for various $m$. At $m\to\infty$, the curve is given by the line $r=2(\sqrt{2}-1)$.}
    \label{fig:BTZ-r-vs-x}
\end{figure}

Consider a boundary region $R$ of length $R$. Let us now punch a hole $H$ of length $(1-r)R$ between two boundary regions $R_1$ and $R_2$ of length $rR/2$ each, similar to as described in section \ref{sec: review}. For bulk reconstruction, we need the entropy of the connected wedge $S_\rm{conn.}$ to be smaller than that of the disconnected wedge $S_\rm{disc.}$,
\begin{equation}\label{eq:BTZ_entropy_compare}
    \begin{split}
        & S_\rm{disc.} \geq S_\rm{conn.}\\
        & \implies S\left(\frac{rR}{2}\right)+ S\left(\frac{rR}{2}\right) \geq S(R) + S\left((1-r)R\right)
        \end{split}
\end{equation}
The entanglement entropy of the disconnected wedge is given by
\begin{equation}
\begin{aligned}
    S_\rm{disc.}&=\frac{2c}{3}\log{\l(\frac{\sinh{(\pi TrR/2)}}{\pi T\epsilon}\r)}\\
    &=\frac{2c}{3}\log{\l(\sinh{(\pi\c r/2)}\r)}-\frac{2c}{3}\log{(\pi T\epsilon)},
\end{aligned}
\end{equation}
where we have defined $\gamma=R T$. For the connected wedge we have,
\begin{equation}
\begin{aligned}
    S_\rm{conn.}&=\frac{c}{3}\log{\l(\frac{\sinh{(\pi T R)}}{\pi T\epsilon}\r)}+\frac{c}{3}\log{\l(\frac{\sinh{(\pi T(1-r)R)}}{\pi T\epsilon}\r)}\\
    &=\frac{c}{3}\log{\l(\sinh{(\pi\c)}\sinh{(\pi\c(1-r))}\r)}-\frac{2c}{3}\log{\l(\pi T\epsilon\r)}. 
\end{aligned}
\end{equation}
In order to find the critical $r$, we have to evaluate the equality in \ref{eq:BTZ_entropy_compare}, which casts itself into the following,
\begin{equation}\label{eq:btz-level-1}
 \sinh^2{\l(\pi\c r/2\r)}=\sinh{(\pi\c)}\sinh{(\pi\c(1-r))},
\end{equation}
whose solution is given by
\begin{equation}
    r=\frac{2}{x}\sinh^{-1}{\l(\sqrt{\frac{(-3 + 2 \sqrt{2} \cosh{x}) \sinh^2{x}}{-5 + 4 \cosh{2x}}}\r)},
\end{equation}
where we have defined $x=\pi \gamma$. Although, this looks different in appearance from the solution given in \cite{Ageev:2022awq}, but they are actually the same. See the purple curve in figure \ref{fig:BTZ-r-vs-x}, showing the temperature dependence of $r$ at level 1.
But what about higher levels? At an arbitrary level $m$, the equality $S_\rm{conn.}=S_\rm{disc.}$ \ref{eq:btz-level-1} takes the form,
\begin{align}\label{eq:btz-level-m}
  \sinh^{2^m}{\l(x(r/2)^m \r)}&=\sinh{x}\nn\\
  &\times\prod_{j=1}^m\sinh^{2^{(j-1)}}{\l(x(1-r)(r/2)^{j-1}\r)}
\end{align}
We plot the relation between $r$ and $x$ for various $m$ in figure \ref{fig:BTZ-r-vs-x} and find a qualitative match with figure \ref{fig:r-vs-Sb-diff-m}, obtained in the case of vacuum-AdS approximation with the bulk entanglement entropy term $S_b$\cite{Bao:2022tgv}.

\begin{figure}[h!]
    \centering
    \includegraphics[scale=0.6]{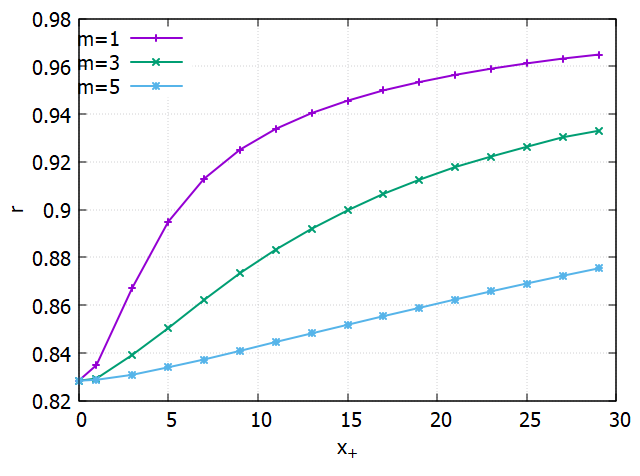}
    \caption{$r$ vs $x_+$ for various $m$ for a rotating BTZ blackhole keeping $x_-=0.01$. We observe a similar qualitative trend as in the case of non-rotating BTZ blackhole in figure \ref{fig:BTZ-r-vs-x}.}
    \label{fig:BTZ-r-vs-x-rotate}
\end{figure}

\begin{figure}[h!]
    \centering
    \includegraphics[scale=0.8]{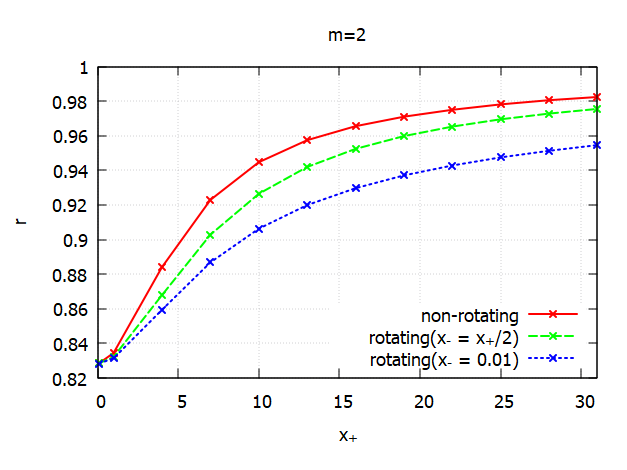}
    \caption{We compare the effect of rotation on $r$ with the non-rotating case at level $m=2$ as one varies $x_+$.}
    \label{fig:BTZ-r-vs-x-rotate-m2}
\end{figure}

Now, consider the case of a rotating BTZ black hole, whose metric is given by \cite{Banados:1992wn}
\begin{equation}
\begin{split}
    ds^{2}=& -\frac{\left(r^{2}-r_{+}^{2}\right)\left(r^{2}-r_{-}^{2}\right)}{r^{2}} d t^{2}+\frac{r^{2}}{\left(r^{2}-r_{+}^{2}\right)\left(r^{2}-r_{-}^{2}\right)} d r^{2}\\
    & +r^{2}\left(d \phi-\frac{r_{+} r_{-}}{2 r^{2}} d t\right)^{2},
\end{split}
\end{equation}
where, due to rotation, now we have two horizons $r_{\pm}$, which are related to the temperature $T_{\pm}$ of left and right moving modes of the dual CFT, defined by
\begin{equation}
    T_{+}=\frac{r_{+}+r_{-}}{2 \pi}, \quad T_{-}=\frac{r_{+}-r_{-}}{2 \pi},\,\,\,\,\,\,\,\,T=\frac{r_{+}^{2}-r_{-}^{2}}{2 \pi r_{+}}.
\end{equation}
The entanglement entropy of a subregion of length $l$ is given by
\begin{equation}
    S(l)=\frac{c}{6} \log \left(\frac{\sinh \left(T_+\pi l\right)}{T_+\pi \epsilon}\right) +\frac{c}{6} \log \left(\frac{\sinh \left(T_-\pi l\right)}{T_-\pi \epsilon}\right).
\end{equation}
Following a similar partitioning of the boundary subregion into two equal disjoint regions $R_1$ and $R_2$ with a hole $H$ between them, as done for the case of non-rotating blackhole, following this up to level $m$, and equating $S_\rm{conn.}$ with $S_\rm{disc.}$ yields
\begin{equation}
        \begin{aligned}
        &\sinh^{2^m}{\l(x_-(r/2)^m \r)}\sinh^{2^m}{\l(x_+(r/2)^m \r)}\\
        &=\sinh{(x_-)}\sinh{(x_+)}\times\nn\\
        &\times\prod_{j=1}^m\sinh^{2^{(j-1)}}{\l(x_-(1-r)(r/2)^{j-1}\r)}\times\nn\\
        &~~~~~~~~\times\sinh^{2^{(j-1)}}{\l(x_+(1-r)(r/2)^{j-1}\r)},
    \end{aligned}
\end{equation}
where we have defined $x_-=\pi RT_-$ and $x_+=\pi RT_+$. We plot a variation of $r$ with $x_+$ for various $m$, keeping $x_-$ fixed, in figure \ref{fig:BTZ-r-vs-x-rotate}, which is in qualitative agreement with the non-rotating case. Next, we fix $m$ and plot the variation of $r$ with $x_+$ for various $x_-$ and contrast with the non-rotating case in figure \ref{fig:BTZ-r-vs-x-rotate-m2}.

One lesson from this exercise is that as long as the value of $r$ favors the connected phase and our minimal surfaces are far from the horizon, uberholographic code properties hold as usual. Moreover, we also established that the results obtained numerically for the BTZ black hole background in $AdS_3$ matches with those obtained in \cite{Bao:2022tgv} using reconstruction wedge arguments in the vacuum-$AdS_3$ limit. We use this evidence to argue that the reconstruction wedged based arguments in the vacuum-AdS metric approximation suffice. In section \ref{sec:d_uber}, we will use the reconstruction wedge method to study uberholography in the presence of bulk entropy in the vacuum-$AdS_d$ limit, neglecting backreaction.

\subsection{The dustball geometry}
This subsection is an extension of the results for BTZ geometry and an uninterested reader may skip to section \ref{sec:d_uber}. The dustball geometry describes an  Oppenheimer-Snyder collapse\cite{Oppenheimer1939} of a spherical ball of pressureless dust of uniform density in $2+1$-dimensional AdS. We take a symmetric time-slice of this geometry. The metric for this geometry can be divided into two parts, each applicable to two different regions. Outside the dustball, the metric is given by a BTZ blackhole metric and inside the dustball, the metric is defined by the FRW universe, with appropriately matched boundary conditions for a continuous metric on the boundary of the dustball. Let us take the $t=0$ slice and consider a dustball of radius $R_d$, we have the following metric outside the dustball,
\begin{equation}
    ds_{out}^2 = \frac{L_{AdS}^2}{r^2-r_h^2} dr^2 + r^2 d\phi^2, \quad r>R_d
\end{equation}
whereas, the metric inside the dustball can be written as
\begin{equation}
    ds_{in}^2 = \frac{1}{1+\frac{r^2}{a_0^2}} dr^2 + r^2 d\phi^2, \quad r<R_d
\end{equation}
where $a_0$ is the scale factor of the interior FRW universe at maximum size.

In the limit where the dustball radius is small, i.e, $R_d \sim r_h$ and our minimal surfaces are far from the dustball, the anaylsis of the BTZ blackhole holds true for the dustball as well. The striking difference between a dustball and blackhole comes when the minimal surfaces are closer to the blackhole horizon, as the minimal surface probes have a horizon avoiding behaviour\cite{Hubeny:2012ry}. 

In the other limit, where the dustball radius is large, i.e., $R_d \gg r_h \gg L_{AdS}$, the connected wedge is more complicated, and we do not give a full discourse here. However, it is worth mentioning that the calculation of disconnected wedge becomes extremely simple in this case. Let us assume that at $t=0$, the dustball has a small energy density $\rho_0$, small in the sense,
\begin{equation}
    \eta =8\pi G_N\rho_0 L^2_{AdS}\ll 1. 
\end{equation}
The Friedmann equation gives
\begin{equation}
    \frac{L_{AdS}^2}{a_0^2}=1-\eta
\end{equation}
and the continuity of metric at $r=R_d$ gives
\begin{equation}
    L_{AdS}^2+r_h^2=\eta R_d^2,
\end{equation}
which together implies that outside the dustball, the metric is near-vacuum to leading order \cite{Akers:2019wxj}
\begin{equation}
    ds_{out}^2=\frac{L_{AdS}^2}{r^2+L_{AdS}^2}(1+\mathcal{O}(\eta^2))dr^2+ r^2d\phi^2, \quad r> R_d
\end{equation}
The minimal surface of the disconnected wedge is the nearly the same as that of the vacuum at leading order corrected by $\eta$-dependent term(s). 

For the connected wedge, at large enough $m$, a large contribution to the connected minimal surface comes from near-boundary curves, as the the size of the holes get smaller with every iteration. These contributions beyond a certain level $m_0$ of hole punching can be computed in the near-vacuum approximation, similar to the case of disconnected wedges. However, for small $m$, this approximation does not hold. Therefore, we do not expect any drastic changes that could alter the physical consequences. As the dustball geometry, in a crude sense, is an intermediate geometry between the blackhole and vacuum $AdS$, it is therefore reasonable to expect that the code properties of uberholography hold in this geometry, albeit, with a complicated expression for the critical $r$.


\section{Uberholography in higher dimensions}\label{sec:d_uber}
Uberholography in higher dimensions was first studied in \cite{Bao:2022vxc} for $AdS_4/CFT_3$ where the authors considered the shape of a Sierpinski triangle as the boundary CFT and a no-go limit was put on its generalization to Sierpinski gasket in $AdS_5/CFT_4$. In another work \cite{Ageev:2022awq}, the author considered uberholohraphy for higher dimensional slicings of Cantor-set-like erasures on one chosen direction in the boundary. It is not a true $d$-dimensional fractal but a $1$-dimensional fractal attached with a $(d-1)$-dimensional space, i.e, $\mathbb{R}\times M_{d-1}$, where the fractal is on $\mathbb{R}$ and $M_{d-1}$ is the space spanned in other directions. Some interesting properties of this (potential) uberholographic code were discussed, which we will review here.

\subsection{An infinite strip}
\begin{figure}[h!]
    \centering
    \includegraphics[scale=0.8]{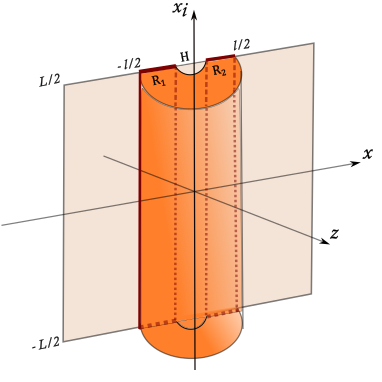}
    \caption{Division of the strip of length $L$ and width $l$ into three narrow strips $R_1, R_2$, and $H$ of widths $rl/2, rl/2$, and $(1-r)l/2$ respectively. We have shown the entanglement wedge for the connected phase only. }
    \label{fig:strip}
\end{figure}

\begin{figure}[h!]
    \centering
    \includegraphics[scale=0.6]{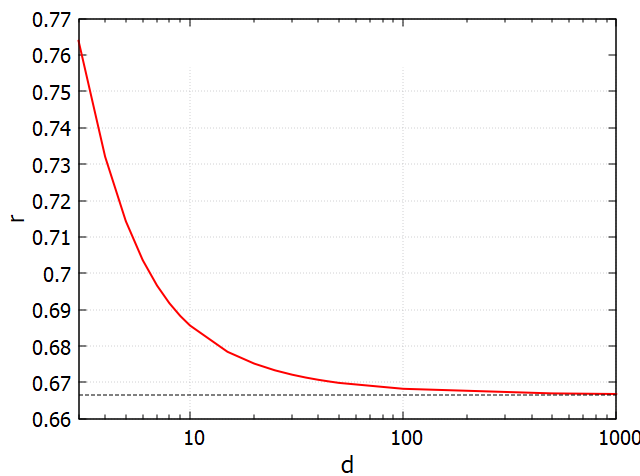}
    \caption{Plot showing the variation of $r$ with increasing $d$: $r$ approaches $2/3$ as $d\to\infty$. }
    \label{fig:ageev_r-vs-d}
\end{figure}

Consider an infinite strip of width $l$ in a constant time-slice of the flat-space boundary $CFT_d$ of an $AdS_{d+1}$ bulk. The entanglement entropy of this strip is given by the Ryu-Takayanagi formula \cite{Ryu:2006ef}
\begin{equation}\label{RT-formula-infinite-strip-any-d}
\begin{aligned}
    S(l)&=\frac{1}{4\pi G^{d+1}} \Bigg[\frac{2L_{AdS}^{d-1}}{d-2}\left(\frac{L}{\epsilon}\right)^{d-2} \\
    & -\frac{2^{d-1}\pi^{{\frac{d-1}{2}}}L_{AdS}^{d-1}}{d-2}\left(\frac{\Gamma\left(\frac{d}{2d-2}\right)}{\Gamma\left(\frac{1}{2d-2}\right)}\right)^{d-1}\left(\frac{L}{l}\right)^{d-2}\Bigg],
\end{aligned}
\end{equation}
where $L$ is the length of the strip ($L\rightarrow\infty$). Notice that the first term is a cutoff-dependent divergent term, however, it does not depend on the width $l$ of the strip. The second term is finite (for large but finite $L$) and universal, which depends on the strip width $l$ but not on the cutoff. So the dependence of entanglement entropy on the strip width $l$ only comes from the second term. In other words, the leading order divergent contribution to the entanglement entropy does not depend on the width of the strip. Thus, a comparison between connected and disconnected phases of the entanglement wedge disregards the leading order divergent term \cite{Ben-Ami:2014gsa} when they appear the same number of times in both cases. This property was exploited in \cite{Ageev:2022awq} for a level-1 comparison. If we divide the infinite strip of width $l$ into three narrower infinite strips of width $rl/2$, $(1-r)l$ and $rl/2$ respectively, as illustrated in figure \ref{fig:strip}, the entropy of the connected phase is given by
\begin{equation}
    S_\rm{conn.} =S(l)+S((1-r)l)
\end{equation}
and that of the disconnected phase is given by
\begin{equation}
    S_\rm{disc.}=2S(rl/2).
\end{equation}
Thus the critical value of $r$ at which connected-disconnected phase transition occurs is given by solving for $r$ when $S_\rm{conn.}=S_\rm{disc.}$,
\begin{align}\label{no_Sb}
    &2\l(\frac{2L}{rl}\r)^{d-2}=\l(\frac{L}{l}\r)^{d-2}+\l(\frac{L}{(1-r)l}\r)^{d-2}\nn\\
    \implies& 2^{d-1}r^2\l(\frac{1}{r}\r)^d-(1-r)^2\l(\frac{1}{1-r}\r)^d-1=0,
\end{align}
    which is equation (3.3) of \cite{Ageev:2022awq}. It easily follows from here that for
\begin{equation}
        d=3:\quad r=3-\sqrt{5}\approx0.763932,
\end{equation}
\begin{equation}
        d=4:\quad r=-1+\sqrt{3}\approx 0.732051.
\end{equation}
and $r\rightarrow 2/3$ for $d\rightarrow\infty$ as shown in figure \ref{fig:ageev_r-vs-d}. Interestingly, the value $r$ saturates at $2/3$, indicating that generalized-Cantor-set-like erasures in any space-time dimensions, the boundary support of the uberholographic code can only approach a zero Lebesgue-measure.
Extending the above calculation to an arbitrary higher-level $m$, we have
\begin{align}\label{no_Sb_m}
    2^m\l(\frac{2}{r}\r)^{m(d-2)}&=1\nn\\
    &+\sum_{j=1}^m2^{j-1}(1-r)^{2-d}\l(\frac{2}{r}\r)^{(d-2)(j-1)}.
\end{align}
The value of $r$ is independent of $m$ similar to the case of Cantor-set-like erasure in a constant-time-slice of $AdS_3/CFT_2$. We think this is a remnant of the nature of the fractal being Cantor-set-like.

\subsubsection{Reconstruction wedges}
\begin{figure}[h!]
    \centering
    \includegraphics[width=\linewidth]{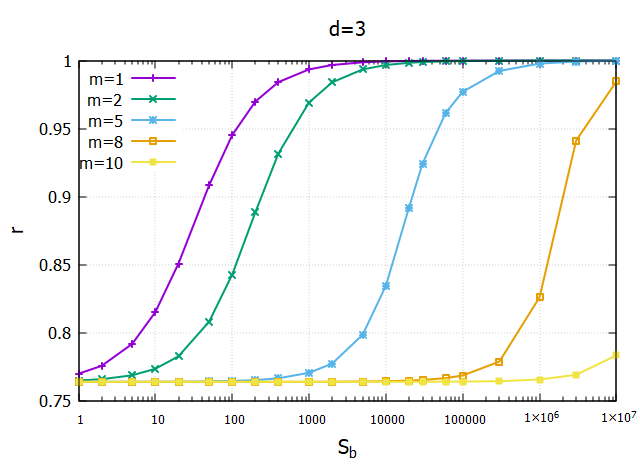}
    \caption{Variation of $r$ with $S_b$ for various $m$ in $AdS_4/CFT_3$.}
    \label{fig:m_d=3}
\end{figure}

Let us now consider adding a bulk entropy $S_b$ in the center of the bulk (we neglect back reaction, as done in \cite{Bao:2022tgv}). At level-1, the entanglement entropy of the connected phase now includes the contribution $S_b$, which can be written as
\begin{equation}
    S_\rm{conn.} =S(l)+S((1-r)l)+S_b
\end{equation}
As in the case of $AdS_3/CFT_2$, the degeneracy of the value of $r$ is broken and it becomes a function of both $m$ and $S_b$. Now the level 1 equation, Eq. (\ref{no_Sb}), changes to
\begin{equation}
     2^{d-1}r^2\l(\frac{1}{r}\r)^d-(1-r)^2\l(\frac{1}{1-r}\r)^d-1=-F_d\cdot S_b,
\end{equation}
where
\begin{align*}
    F_d=\frac{d-2}{2^{d-1}\pi^{d-1}}\l(\frac{\G\l(\frac{1}{2d-2}\r)}{\G\l(\frac{d}{2d-1}\r)}\r)^{d-1}\frac{4G}{L_{AdS}^{d-1}l_\perp^{d-2}}.
\end{align*}
For arbitrary level $m$, we get the analogue of Eq. (\ref{no_Sb_m}),
\begin{align}
    &2^m\l(\frac{2}{r}\r)^{m(d-2)}\nn\\
    -&\sum_{j=1}^m2^{j-1}(1-r)^{2-d}\l(\frac{2}{r}\r)^{(d-2)(j-1)}-1\nn\\
    &~~~~~~~~~~~~~~~~~~~~~~~~~~~~~~~~~~~~~~~~~~~~=-F_d\cdot S_b.
\end{align}
We plot the dependence of $r$ on $S_b$ for various $m$ in $d=3$ below (figure \ref{fig:m_d=3}), for instance.

We observe a qualitative similarity of trend in comparison to $AdS_3/CFT_2$ (see figure \ref{fig:r-vs-Sb-diff-m}). The lesson we learn here is that the uberholographic code for Cantor-set-like fractal boundary erasures in $AdS_3/CFT_2$ discovered in \cite{Pastawski:2016qrs} is the simplest case of a family of uberholographic codes with Cantor-set-like fractal-like boundary erasures in $AdS_{d+1}/CFT_d$, albeit the higher dimensional cases are not really fractals in the true sense (therefore, fractal-like). The Sierpinski triangle-like uberholographic code found in \cite{Bao:2022tgv}, however, do not belong to this family. Such codes are robust against the breakdown of entanglement wedge in the presence of highly entropic mixed states in the bulk.

\subsubsection{Code distance}
Now we will characterize the code distance of the code built from infinitely long, straight strip in $AdS_{d+1}/CFT_{d}$, corresponding to the scenario when $S_b=0$. If we start with a strip of length $L$ and width $l$, after punching a hole at level $m$, we will be left with strips of width $(r/2)^ml$. The thinnest strip we can have is of width $\epsilon$, the short-distance cut-off. Then we have $2^m$ strips, each of width $\epsilon$, and hence the code distance is given by (see Eq. (\ref{codedist}))
\begin{align}
    d(\mathcal{A}_X)\leq2^m=\l(\frac{l}{\epsilon}\r)^\a,
\end{align}
where
\begin{align}\label{eq:alpha}
    \a=\frac{\log{2}}{\log{2/r}}=\begin{cases}
        0.720\quad(d=3)\\
        0.689\quad(d=4)\\
        \vdots\\
        0.631\quad(d\to\infty)
    \end{cases}
    ,
\end{align}
which is the formula for Hausdorff dimension for a generalized Cantor-set fractal in one dimension. Thus, the value of $\alpha$ reflects the fact that in our straight Cantor-set slicing in higher dimensions, the `fractalness' is only along one dimension.
Recall that for the case of Cantor-set like fractal in $AdS_3/CFT_2$, the value was $\alpha=0.786$\cite{Pastawski:2016qrs} given by the same expression \ref{eq:alpha}. For the Sierpinski triangle code in $AdS_4/CFT_3$, the value of $\alpha$ was not related to expression \ref{eq:alpha} and was computed to be $\alpha=0.7925$\cite{Bao:2022tgv}, exactly half of the Hausdorff dimension of the Sierpinski triangle fractal.

\subsection{The role of infinitely long, straight geometry}
We discuss the crucial roles played by `infinite' length and `straight' shape in the following subsections. We will relax these properties one at a time and study their effect on the code.

\subsubsection{A finite long straight strip}
The entanglement entropy of a finitely long straight strip of length $L$ and width $l$ upto leading order, is given by \cite{Fonda:2014cca}
\begin{equation}\label{eq:finite-L-entropy}
    S(l)=\frac{2c}{3}\left(\frac{L+l}{\epsilon} + \cdots \right)
\end{equation}
where the $\cdots$ include a sub-leading logarithmic term.

Consider the expression \ref{eq:finite-L-entropy}. Now, we want to solve for $r$ where $S_{\rm{conn.}}= S_{\rm{disc.}}$ such that $0<r<1$. At level 1,
\begin{equation}
    \begin{split}
        & S_{\rm{conn.}}= S_{\rm{disc.}}\\
        &\implies S(l)+S((1-r)l)=2S(rl/2)\\
        & \implies 1+(1-r)=r\\
        & \implies r=1 \quad(\text{not allowed})
    \end{split}
\end{equation}
At an arbitrary level $m$, we have the condition
\begin{align}
    1+(1-r)\sum_{j=0}^{m-1}r^j = r^m, \implies r^m= 1\quad(\text{not allowed}).
\end{align}
The critical value $r=1$ implies that support on the entire boundary slice is required for bulk reconstruction and erasures on the boundary are not correctable. Thus, the given geometry does not have any code properties.

\subsubsection{An annulus stretched to a long (infinite) strip}
The entanglement entropy of an annular region between radii $R_1$ and $R_2$ (say, $R_2>R_1$) is given by\cite{Han_2022},
\begin{equation}
    S(R_1, R_2)= \frac{c}{6}\left[\frac{2\pi R_2}{\epsilon_2}+\frac{2\pi R_1}{\epsilon_1} -4\pi \frac{R_2^2+R_1^2}{R_2^2-R_1^2}\right]
\end{equation}
where $c$ is the central charge. Note that the cutoffs $\epsilon_1$ and $\epsilon_2$ are in general different. WLOG, we can write $\epsilon_2=a\epsilon_1$. For the sake of the simplicity we take $a=1$(i.e, $\epsilon_1=\epsilon_2$, which is not true and the value of $a$ must be inserted back to correctly evaluate the entanglement entropy. But it suffices to not worry about $a$ to demonstrate our argument. Henceforth, we drop the subscript from the cutoffs and call it $\epsilon$). One can take the limit $2\pi R_1= L\rightarrow \infty$ and $2\pi R_2=(L+l)$, which takes us to the infinitely long strip (without edges along width). The entanglement entropy is given by (see Appendix \ref{appb} for another variant of this derivation),
\begin{align}
S(L,l)&= \frac{c}{6}\left[\frac{(L+l)}{\epsilon}+\frac{ L}{\epsilon} -4\pi \frac{(L+l)^2+L^2}{(L+l)^2-L^2}\right]\nn\\
&= \frac{c}{6}\left[\frac{2 L}{\epsilon}+ \frac{ l}{\epsilon} -4 \pi \frac{2+2l/L+l^2/L^2}{(l/L)(1+2l/L)}\right]
\end{align}
Keeping the sub-leading terms in $l/L$ upto linear order, it has the following form
\begin{equation}\label{eq:annulus-RT-formula}
     S(l)=\frac{c}{6}\left[\frac{2L}{\epsilon}+\frac{l}{\epsilon}-4\pi\frac{L}{l}+\cdots\right]
\end{equation}
where $\cdots$ refer to terms higher order in $l/L$.
To leading order in $\frac{1}{\epsilon}$, this is similar to equation \ref{eq:finite-L-entropy}, and therefore, code properties do not exist. Let us now include the finite term $L/l$ to see whether the inclusion of this term gives rise to code properties. The functional form of the entanglement entropy $S(l)$ is given by,
\begin{equation}
    S(l)=\frac{Al}{\epsilon}+BL\bigg(\frac{1}{\epsilon}-\frac{1}{l}\bigg)
\end{equation}
At level 1, we have
\begin{equation}
\begin{aligned}
    &2S(rl/2)=S(l)+S((1-r)l)\\
    \implies& \frac{Arl}{\epsilon}-\frac{4BL}{rl}=\frac{Al}{\epsilon}-\frac{BL}{l}+\frac{A(1-r)l}{\epsilon}-\frac{BL}{(1-r)l}\\
    \implies & r=\frac{(D-\epsilon)\pm\sqrt{5\epsilon^2+\epsilon D}}{D+\epsilon}\quad\text{where}\quad D\equiv \frac{2Al^2}{BL}
\end{aligned}
\end{equation}
We are only interested in the limit where $\epsilon\ll D$. Recall that $D=\frac{2Al^2}{BL} \sim \frac{l^2}{L}$. Since the strip length $L\rightarrow\infty$ and the short-distance cutoff $\epsilon\rightarrow 0$, we have $r\rightarrow 1$. Thus, code properties are \emph{practically} non-existent.

\subsubsection{Remarks}
\begin{figure}[h!]
    \centering
    \begin{subfigure}{0.35\textwidth}
        \centering
        \includegraphics[width=\textwidth]{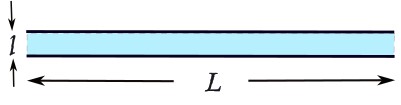}
        \caption{Infinite straight belt of width $l$ and length $L\to\infty$.}
        \label{fig:sub1}
    \end{subfigure}
    \hfill
    \begin{subfigure}{0.35\textwidth}
        \centering
        \includegraphics[width=\textwidth]{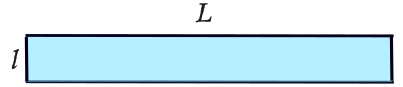}
        \caption{Long rectangular strip of length $L$ and width $l$.}
        \label{fig:sub2}
    \end{subfigure}
    \hfill
    \begin{subfigure}{0.4\textwidth}
        \centering
        \includegraphics[width=\textwidth]{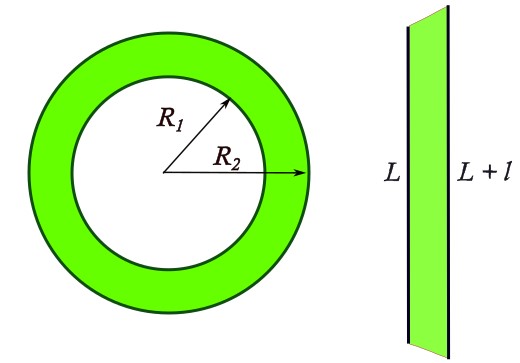}
        \caption{Annulus of inner and outer radii $R_1=L/2\pi$ and $R_2=(L+l)/2\pi$.}
        \label{fig:sub3annulas}
    \end{subfigure}
    \caption{The three cases considered in the text.}
    \label{fig:3_cases}
\end{figure}

As has been pointed out in \cite{Bombelli:1986rw, Srednicki:1993im}, the entanglement entropy of a subregion $R$ in a spatial time slice of a quantum field theory in $d$-dimensions, upto leading order is given by
\begin{equation}
    S(R)= \gamma \frac{\text{Area}(\partial R)}{\epsilon^{d-2}} + \cdots
\end{equation}
where $\epsilon$ is the short distance cutoff and $\cdots$ correspond to sub-leading terms. In the above three cases, this holds readily,
\begin{itemize}
    \item the infinite, straight belt of length $L\to\infty$ and width $l$ has only the length $L$ on two sides contributing to its boundary;
    \item the finite, long strip of length $L$ and width $l$ has the length $L$ as well as the width $l$, on both sides respectively contributing to its boundary;
    \item the annulus of radii $L/2\pi$ and $(L+l)/2\pi$, stretched to the limit $L\rightarrow\infty$, to make an infinite strip, although has only the length on two sides contributing to the boundary, but these sides are of unequal lengths, i.e, one side (corresponding to inner radii) has length $L$ whereas the other side (corresponding to outer radii) has a length $(L+l)$.
\end{itemize}
We illustrate the three cases in figure \ref{fig:3_cases}. Since the error correction properties hold only in the true $L\rightarrow\infty$ limit, such code properties have almost no practical value. Thus, only the true Cantor-set fractal code in $AdS_3/CFT_2$ has practical significance at finite sizes, whereas all Cantor-set fractal-like codes in $AdS_{d+1}/CFT_{d}$, $(d\geq3)$ are asymptotic codes only in the infinite size limit. In contract, the uberholographic Sierpinski triangle\cite{Bao:2022vxc} in $AdS_4/CFT_3$ retains code properties at all sizes and therefore has a more practical significance.

\section{Comments on uberholography}
\label{sec:comments}
In this section, we comment on uberholography from the perspective of two recent developments in holography. The first comment sheds light on the operator algebraic properties of uberholography and the complexity transfer between connected and disconnected phases, along the lines of \cite{Engelhardt:2023xer}. The second comment puts light on uberholography with respect to the new paradigm of bulk reconstruction where it is seen as `one-shot state merging'\cite{Akers:2020pmf,Akers:2023fqr}. This picture introduces leading order corrections to the QES prescription.

\subsection{Operator algebra perspective to uberholography}
The idea of complexity transfer at the exchange of dominance between connected and disconnected QES was given by \cite{Engelhardt:2023xer}.
The two central claims of their work are the following:
\begin{lemma}\label{lemma:netta-hong-algebra}
Let $\ket{\psi_{R_1 R_2}}$ be a pure bipartite state on two copies of the static cylinder, and let $\mathcal{W}_{R_1\cup R_2}$ be its semiclassical dual description in the large-N limit. Then, $\mathcal{W}_{R_1\cup R_2}$ is
\begin{itemize}
    \item disconnected, iff $\mathcal{A}_{R_1}$ and $\mathcal{A}_{R_2}$ are type I;
    \item classically connected, iff $\mathcal{A}_{R_1}$ and $\mathcal{A}_{R_2}$ are type III$_1$,
\end{itemize}
where $\mathcal{A}_{R_i}$ is the boundary algebra on $R_i$.
\end{lemma}

Generally, the systems $R_1$ and $R_2$ are constant time Cauchy slices of CFT, having the full boundary algebra $\mathcal{A}_{R_1}$ and $\mathcal{A}_{R_2}$ respectively. However, in our context of uberholography, we are dealing with boundary subregions of the same CFT. So, when considering two boundary subregions $R_1$ and $R_2$, we implicitly mean that we are talking about those regions and their algebra after canonical purification. The classical (dis)connectedness of canonically purified $R_1$ and $R_2$ are by construction related to the (dis)connected phase of entanglement wedge of boundary subregions.

\begin{lemma}\label{lemma:netta-hong-complex}
Switchovers in the dominant QES generically originate in the dynamical transfer of a type III$_1$ factor consisting of operators of high complexity. 
\end{lemma}

\begin{figure}[h!]
    \centering
    \begin{subfigure}{0.4\textwidth}
        \centering
        \includegraphics[width=0.5\textwidth]{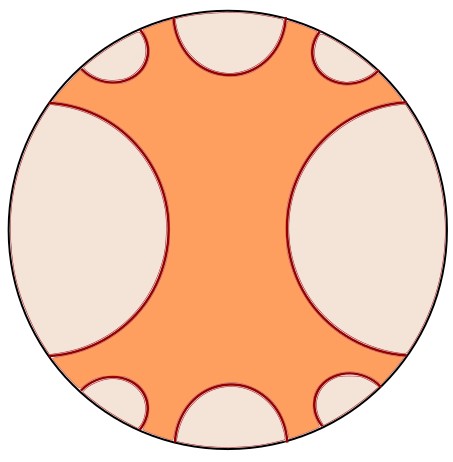}
        \caption{The connected phase (shaded) of $R$. This region is also a min-wedge corresponding to boundary region $R$.}
        \label{fig:sub1con}
    \end{subfigure}
    
    \begin{subfigure}{0.4\textwidth}
        \centering
        \includegraphics[width=0.5\textwidth]{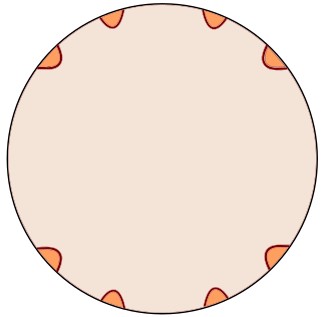}
        \caption{The disconnected phase (shaded) of $R$. This region is also a max-wedge corresponding to boundary region $R$.}
        \label{fig:sub2disc}
    \end{subfigure}
    \caption{The connected and disconnected phases. Equivalently, they also correspond to the max and min wedges corresponding to boundary region $R$ at $r=r_c$.}
    \label{fig:conn_disc}
\end{figure}

In our context, the boundary subregion $R$ is an union of disjoint smaller subregions $\bigcup_i R_i$ of equal sizes, but non-uniformly separated (see fig \ref{fig:conn_disc}).

For simplicity, consider the case $S_b=0$ where $r_c$ is independent of $m$. The above lemmas \ref{lemma:netta-hong-algebra},\ref{lemma:netta-hong-complex} can be applied to any two `bipartition' of the multipartite state (with canonical purification). For the disconnected phase, since $r<r_c$, no `bipartite' combination of subregions from the set of subregions $\{R_i\}$ can be classically connected.

We may now say that the disconnected phase admits a factorization of Hilbert space $\mathcal{H}_R=\otimes \mathcal{H}_{R_i}$ and any local operator $\hat{O}$ can be reconstructed with sub-exponential complexity. From an operator algebra perspective\cite{Witten:2018zxz,Leutheusser:2022bgi}, in the disconnected phase, the subalgebra $\mathcal{A}_{R_i}$ associated with each subregion $R_i$ is \emph{type I}. The entanglement wedges $\mathcal{E}[R_i]$ of each $R_i$ are \emph{classically disconnected} from each other (see figure \ref{fig:sub2disc}). At $r>r_c$, the dominance of the of entanglement wedges are exchanged. The entanglement wedge associated with the connected phase is \emph{classically connected} and therefore the algebra on $R$ is not \emph{type I}. We conjecture that the subalgebra of $R$ is \emph{type} $III_1$ (this is a little subtle to prove, because not every pair of boundary subregion are classically connected and therefore the construction used in \cite{Engelhardt:2023xer} cannot be used directly). This is facilitated by the transfer of complex factor to the subalgebra at the transition and there doesn't exist any simple way to reconstruct all operators in $\mathcal{W}_R$.

\subsection{One-shot uberholography}
In our work using the entanglement wedge and reconstruction wedge arguments, the phase transition over the exchange of dominance between connected and disconnected entanglement wedges is a discontinuous one and is of the nature \emph{all or nothing.}, i.e, we can, \emph{in principle}, fully reconstruct any operator in the center of the bulk in the connected phase, whereas no such operator can be reconstructed in the disconnected phase.

It was pointed out in \cite{Akers:2020pmf,Akers:2023fqr} that the von Neumann entropy (and, the QES pescription) has no operational significance when you have only one copy of the bulk, unless the subregions satisfy the \emph{all or nothing} recovery for the given quantum state, i.e, in those cases, the von Neumann entropy is equal to two other operational quantities, namely, \emph{max-entanglement entropy} and \emph{min-entanglement entropy}, which we define below. Any system with a flat Renyi-entropy spectrum satisfies this condition. Just like the reconstruction wedge is the bulk wedge corresponding to von Neumann entropy, their corresponding wedges are \emph{max-wedge} and \emph{min-wedge} respectively (see \cite{Bao:2024hwy} for a classification of wedges).
We first define the smooth conditional max-entropy $H_{\text{max}}^\epsilon(AB|B)_\rho$ as
\begin{equation}
    \label{eq:max-entropy}
    H_{\text{max}}^\epsilon(AB|B)_{\rho}=\inf_{\tilde{\rho} \in \mathcal{B}^\epsilon(\rho)} \sup_{\sigma_B} \ln\left(\tr_A\left[\sqrt{\sigma_B^{1/2} \tilde{\rho}_{AB} \sigma_B^{1/2}}\right]\right)^2,
\end{equation}
where $\mathcal{B}^\epsilon(\rho)$ is the set of density matrices whose purified distance with $\rho_{AB}$ is smaller than $\epsilon$. We get the unconditional max-entanglement entropy taking $B$ to be trivial.

We define the smoothened conditional min-entanglement entropy to be
\begin{equation}
    \label{eq:min-entropy}
    H_{\text{min}}^\epsilon(AB|B)_\rho=\sup_{\tilde{\rho} \in \mathcal{B}^\epsilon(\rho)} \left(-\min_{\sigma_B} \inf \{ \lambda: \tilde{\rho}_{AB} \leq e^{\lambda} \mathbf{1}_A \otimes \sigma_B\} \right).
\end{equation}
They are related to each other and the von Neumann entropy by,
\begin{equation}
\begin{aligned}
\lim_{n \to \infty} \frac{1}{n} H_{\text{max}}^\epsilon(A^n|B^n)_{\psi^{\otimes n}}&=S(A|B)_\psi \\ &=\lim_{n \to \infty} \frac{1}{n} H_{\text{min}}^\epsilon(A^n|B^n)_{\psi^{\otimes n}}
\end{aligned}
\end{equation}
where $n$ is the number of copies of the holographic state.

Operationally, in simple words, given a single copy of the holographic state, the \emph{max-wedge(A)} is the largest bulk region homologous to $A$ within which all operators can be reconstructed. The \emph{min-wedge(A)} is the largest bulk region homologous to $A$ beyond which no operator can be reconstructed.

We note that in our context, these new definitions add extra complications and hazes the picture of bulk reconstruction for the case of min-wedges. In the previous sections, we have seen that uberholography is consistent with the QES picture. Now let us comment from the perspective of max(min)-wedges. There is an important duality relation between max(min)-entanglement entropies. For any pure state $\psi \in \mathcal{H}_A \otimes \mathcal{H}_B \otimes \mathcal{H}_C$,
\begin{equation}
    \label{duality}
H_{\text{min}}^\epsilon(AB|B)_\psi=-H_{\text{max}}^\epsilon(AC|C)_\psi.
\end{equation}

\begin{figure}[h!]
    \centering
    \includegraphics[width=0.25\textwidth]{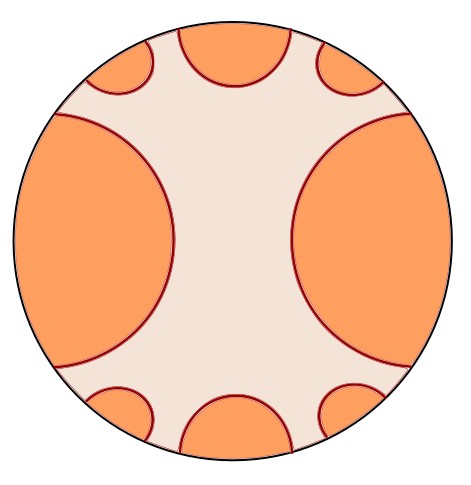}
    \caption{The max-wedge (shaded) corresponding to the erased boundary regions $H=R^c$.}
    \label{fig:holes}
\end{figure}

Recall that the boundary subregion $R$ is a union of disjoint smaller subregions $\bigcup_i R_i$ of equal sizes, whereas the erasure on the boundary is given by the complement $\bar{R}$, which is also a union of disjoint smaller subregions $\bigcup_i H_i$ of unequal sizes (see figures \ref{fig:sub2disc} and \ref{fig:holes}).

The max-wedge is generally contained within the QES entanglement wedge, which is usually contained within the min-wedge. Let us consider the scenario $r\geq r_c$ such that the max-wedge$(H)\in \mathcal{E}[H]$ is disconnected. Moreover, since $\mathcal{E}[R]$ is connected, then by $\mathcal{E}[R]\in$ min-wedge$(R)$, the min-wedge$(R)$ must be connected. The duality in equation \ref{duality} suggests the same. So, there exist some operator(s) deep in the bulk that are reconstructible from the boundary $R$. However, it is not guaranteed that all operators can be reconstructed as the max-wedge$(R)$, may or may not be connected, and in general depends on $r\geq r_c$ as well as the quantum state $\ket{\psi}_R$ on the boundary $R$. But we expect that for sufficiently large $r$, the max-wedge$(R)$ coincides with the entanglement wedge $\mathcal{E}[R]$ for most states.
Now consider the case $r\leq r_c$ where $\mathcal{E}[R]$ is disconnected. Clearly the max-wedge$(R)$ is contained inside and thus disconnected. It is unclear if the min-wedge$(R)$ is connected or disconnected, but we expect it to be disconnected for a sufficiently small $r$ for most states. Therefore, one-shot uberholography changes the picture of a sudden \emph{all or nothing} to a gradual one, i.e, there exists a regime where it is neither all nor nothing, interpolating the cases of all and nothing.

\section{Discussion}
In this revisit of uberholography, we have argued for uberholography as a physical property of some holographic fractal (and fractal-like) geometries, filled up some gaps in the literature and addressed some of questions posed in the discussion of \cite{Bao:2022vxc,Bao:2022tgv}. Incorporating uberholography in the modern paradigms of complexity transfer and one-shot state-merging, we believe this will enable us to shed more light on the fundamental properties of uberholography. During the completion of this work, an attempt to demistify uberholography appeared in \cite{Wang:2024gvz}, where the author proposes a rearrangement of boundary physical qubits to explain the connectedness in an alternative geometry. Although, it is able to explain uberholography, however, it is not well-motivated why such an alternative geometry should exist, or why one geometry should be preferred over another, especially when the structure of locality gets altered.

It is an interesting avenue to consider uberholography in more exotic setups involving cosmic branes and holographic BCFTs. We leave this question for future research. Another interesting direction is to study holographic entropy inequalities\cite{Bao:2015bfa,Hernandez-Cuenca:2023iqh,Czech:2023xed,Bao:2024vmy,Bao:2024azn} in the limit of very large number of boundary-regions, as envisioned in \cite{Czech:2023xed}. While it is easier to deal with equal-sized boundary subregions, fractal geometries are simple yet non-standard setting for such a study.

\begin{acknowledgments}
We thank Ning Bao, Keiichiro Furuya, Dileep Jatkar, Yikun Jiang and Tadashi Takayanagi for useful discussions. J.N. is supported by the NSF under Cooperative Agreement PHY2019786 and Northeastern University.

\end{acknowledgments}

\appendix
\label{appa}
\section{Ryu-Takayanagi's derivation of entanglement entropy of an infinite strip}
Here we review an outline of the derivation of the entanglement entropy of an infinite strip \cite{Ryu:2006ef}. We start with the $AdS_{d+1}$ metric in Poincare coordinates
\begin{equation}
    ds^2=\frac{L_{AdS}^2}{z^2}\left(-dt^2+dz^2+dx^2+\sum_{i=1}^{d-2}dx_i^2\right).
\end{equation}
We want to calculate the area element of the minimal surface for a fixed time slice ($dt=0$) and regulated $x_i$ directions at length $L$,
\begin{align}
    A_S=\{x^\m|x\in[-l/2,l/2],x_i\in[-L/2,L/2]\}.
\end{align}
If we parameterize $z=z(x)$, then the line element becomes
\begin{align}
    ds^2=\frac{L_{AdS}^2}{z^2}\left((1+z'(x)^2)dx^2+\sum_{i=1}^{d-2}dx_i^2\right).
\end{align}
The area of the strip is then computed as follows:
\begin{align}
    \rm{Area}(A_S)&=\int\sqrt{|g|}dxdx_i\nn\\
    &=\left(\prod_{i=1}^{d-2}\int_{-L/2}^{L/2}dx_i\right)\int_{-l/2}^{l/2}dx\frac{L_{AdS}^{d-1}}{z^{d-1}}\sqrt{1+z'^2}\nn\\
    &=L_{AdS}^{d-1}L^{d-2}\int_{-l/2}^{l/2}dx\frac{\sqrt{1+z'^2}}{z^{d-1}}.
\end{align}
Extremizing this area itegral is equivalent to extremizing the following action using the Lagrangian and Hamiltonian formalism:
\begin{align}
    S=\int dx\mathcal L\quad\text{with}\quad\mathcal L=\frac{\sqrt{1+z'^2}}{z^{d-1}}.
\end{align}
The conjugate momentum is given by $P=\frac{\pa\mathcal L}{\pa z'}$ and hence the Hamiltonian is
\begin{align}
    H=Pz'-\mathcal L=-\frac{\sqrt{1-P^2z^{2d-2}}}{z^{d-1}}.
\end{align}
We would like to express $z'$ in terms of $z$. To this end, we perform some manipulations to arrive at
\begin{align}
    z'=\frac{\pa H}{\pa P}=\frac{Pz^{d-1}}{\sqrt{1-P^2z^{2d-2}}}=\frac{\sqrt{1/H^2-z^{2d-2}}}{z^{d-1}}.
\end{align}
If we now set $1/H^2=z_*^{2d-2}$, then we have a turning point at $z=z_*$ and therefore we constrain such that
\begin{align}\label{A8}
    \int_0^{l/2}dx&=\int_0^{z_*}\frac{dz}{z'}=\int_0^{z_*}dz\frac{z^{d-1}}{\sqrt{z_*^{2d-2}-z^{2d-2}}}\nn\\
    \implies\frac{l}{2}&=\int_0^1 duz_*\frac{u^{d-1}}{\sqrt{1-u^{2d-2}}}=\frac{\sqrt{\pi}\Gamma\left(\frac{d}{2d-2}\right)}{\Gamma\left(\frac{1}{2d-2}\right)}z_*,
\end{align}
where we have made the substitution $u=z/z_*$ and used
\begin{align*}
    \int_0^1dxx^{\m-1}(1-x^\la)^{\n-1}=\frac{1}{\la}B\left(\frac{\m}{\la},\n\right)=\frac{\Gamma(\m/\la)\Gamma(\n)}{\la\Gamma(\m/\la+\n)}.
\end{align*}
Going back to the area element, we have
\begin{align}\label{A9}
    \rm{Area}(A_S)&=2L_{AdS}^{d-1}L^{d-2}z_*^{d-1}\times\nn\\
    &\times\bigg[\int_0^{z_*}dz\frac{1}{z^{d-1}\sqrt{z_*^{2d-2}-z^{2d-2}}}\nn\\
    &-\int_0^{\epsilon}dz\frac{1}{z^{d-1}\sqrt{z_*^{2d-2}-z^{2d-2}}}\bigg]\nn\\
    &=\frac{2L_{AdS}^{d-1}}{d-1}\left(\frac{L}{\epsilon}\right)^{d-2}-2IL_{AdS}^{d-1}\left(\frac{L}{z_*}\right)^{d-2}
\end{align}
where $\epsilon$ is the short distance cut-off and 
\begin{align*}
    I\equiv-\frac{\sqrt{\pi}\Gamma\left(\frac{2-d}{2d-2}\right)}{(2d-2)\Gamma\left(\frac{1}{2d-2}\right)}.
\end{align*}
Substituting the form of $z_*$ from Eq. (\ref{A8}) into the second term, (\ref{A9}) can be manipulated to give
\begin{align}
    \rm{Area}(A_S)&=\frac{2L_{AdS}^{d-1}}{d-2}\left(\frac{L}{\epsilon}\right)^{d-2}\nn\\
    &-\frac{2^{d-1}\pi^{\frac{d-1}{2}}}{d-2}\left(\frac{\Gamma\left(\frac{d}{2d-2}\right)}{\Gamma\left(\frac{1}{2d-2}\right)}\right)^{d-1}\l(\frac{L}{l}\r)^{d-2}.
\end{align}

\section{Han-Wen's derivation of entanglement entropy of an infinite strip}
\label{appb}
Here we briefly discuss an alternate method of obtaining the entanglement entropy of an infinite strip~\cite{Han_2022}. This method uses the \textit{additive linear combination (ALC)} proposal for the \textit{partial entanglement entropy (PEE)}. The PEE of any subset $A_i$ of a region $A$ captures the contribution of $A_i$ to the total entanglement entropy $S_A$, and is denoted by $s_A(A_i)$. In $d$ dimensions, it is defined by
\begin{align}
    s_A(A_i)=\int_{A_i}s_A(\bs x)d^{d-1}x,
\end{align}
where $s_A(\bs x)$ is called the \textit{entanglement contour} function for $A$. The ALC proposal goes as follows: If we have a region $A$ and an arbitrary subset $\a$ such that $A$ can be partitioned into three non-overlapping subsets unambiguously, $A=\a_L\cup\a\cup\a_R$, $\a_L(\a_R)$ being subsets to the left(right) of $\a$, then the PEE of $\a$ is given by
\begin{align}\label{ALC}
    s_A(\a)=\frac{1}{2}\l(S_{\a_L\cup\a}+S_{\a_R\cup\a}-S_{\a_L}-S_{\a_R}\r).
\end{align}
The ALC proposal can be used for one dimensional regions, where there is a natural order ("left/right"), or for highly symmetric configurations in higher dimensions where an order can be naturally defined.

We start with the contour function for $(d-1)$ dimensional balls with radius $R$~\cite{Han_2022},
\begin{align}
    s_A(r)=\frac{c}{6}\bigg(\frac{2R}{R^2-r^2}\bigg)^{d-1},
\end{align}
specialized for the case of a disk ($d=3$). Next we consider an annulus $A$, with inner and outer radii $R_1$ and $R_2$ respectively, which we divide into two thinner annuli $A_1$ and $A_2$ by a circle of radius $R$ such that $R_1\leq R\leq R_2$. All cutoff regions are narrow annuli with width $\epsilon$ (see figure \ref{fig:sub3annulas}).
The entanglement entropy of an annulus of inner and outer radii $a$ and $b$ respectively is obtained by adding the entanglement entropies $S_\rm D$ of a disk $\rm D$ of radius $b$ and $S_\rm d$ of a smaller disk $\rm d$ of radius $a$ and subtracting the PEE $s_{\rm D}(\rm d)$:
\begin{align}
    S_\rm{ann.}&=\bigg(\int_0^{b-\epsilon}s_{\rm D}(r)+\int_{0}^{a-\epsilon}s_{\rm d}(r)-\int_0^{a}s_{\rm D}(r)\bigg)2\pi r\,dr\nn\\
    &=\frac{2\pi c}{3}\bigg(\frac{(b-\epsilon)^2}{(2b-\epsilon)\epsilon}+\frac{(a-\epsilon)^2}{(2a-\epsilon)\epsilon}-\frac{2a^2}{b^2-a^2}\bigg)
\end{align}
Using this we can get the entanglement entropies of the annuli $A_1$ and $A_2$:
\begin{align}
    S_{A_1}(R)&=\frac{2\pi c}{3}\bigg(\frac{2R_1^2}{R_1^2-R^2}+\frac{(R-\epsilon)^2}{(2R-\epsilon)\epsilon}+\frac{(R_1-\epsilon)^2}{(2R_1-\epsilon)\epsilon}\bigg)\\
    S_{A_2}(R)&=\frac{2\pi c}{3}\bigg(\frac{2R_2^2}{R^2-R_2^2}+\frac{(R-\epsilon)^2}{(2R-\epsilon)\epsilon}+\frac{(R_2-\epsilon)^2}{(2R_2-\epsilon)\epsilon}\bigg).
\end{align}
Then using the derivative version of the ALC proposal, obtained by applying (\ref{ALC}) to spherical shells (quasi-one-dimensional configuration)~\cite{Han_2022}, leads us to the contour function for our annulus $A$,
\begin{align}
    s_A(r)&=\frac{1}{4\pi r}\pa_R(S_{A_1}-S_{A_2})\bigg\rvert_{R=r}\nn\\
    &=\frac{2\pi c}{3}\bigg(\frac{R_2^2}{(r^2-R_2^2)^2}+\frac{R_1^2}{(R_1^2-r^2)^2}\bigg).
\end{align}
If we set $R_1=R=L/2\pi$, $R_2=R_1+l$, (as before) and $x=R+l/2-r$ we are led to the contour function of a strip, with center at $x=0$, having width $l$ and length $L(\gg l)$:
\begin{align}
    s_\rm{strip}(x)=\frac{2c}{3}&\Bigg[\frac{1+2l/R+l^2/R^2}{\big(\frac{(l/2-x)^2}{R}-(l+2x)\big)^2}\nn\\
    &+\frac{1}{\big(\frac{(l/2-x)^2}{R}+(l-2x)\big)^2}\Bigg].
\end{align}
Note that if we take the limit $L\to\infty$ we recover Eq. (54) of ~\cite{Han_2022}. Integrating $s_\rm{strip}(x)$ from $-l/2+\epsilon$ to $l/2-\epsilon$ and then keeping terms linear in $l/R$ while neglecting terms of order $l^2/R$ and higher, we get
\begin{align}
    S_\rm{strip}&=\frac{c}{3}L\bigg(\frac{1}{\epsilon-l}+\frac{1}{\epsilon}+\frac{2\pi l}{L\epsilon}+\cdots\bigg)\nn\\
    &=\frac{c}{3}L\bigg(\frac{1}{\epsilon}-\frac{1}{l}+\mathcal O(\epsilon)+\frac{2\pi l}{L\epsilon}+\cdots\bigg).
\end{align}
Taking the limit $\frac{l}{L}\rightarrow 0$, we recover the form given by Ryu-Takayanagi \ref{RT-formula-infinite-strip-any-d} for $d=3$.

\bibliography{main}

\end{document}